\documentclass[10pt]{article}
\usepackage[utf8]{inputenc}
\usepackage{multirow} 
\usepackage{graphicx,float,lipsum}
\usepackage{wrapfig}
\usepackage{multicol}
\usepackage{braket}
\usepackage[a4paper,top=2cm,bottom=2cm,left=3cm,right=3cm,marginparwidth=1.75cm]{geometry}
\usepackage{xcolor}
\usepackage{subfigure} 
\usepackage{amsmath}\usepackage[colorlinks=true, allcolors=blue]{hyperref}
\usepackage{graphicx}
\usepackage{amssymb}

\newcommand{\dH}{\mathrm{d}_H}

\title{\textbf{Numerical estimation of the Hausdorff dimension of $D$-random feuilletages}}
\author{\textbf{Alicia Castro}$^{a}$\footnote{Email: \href{mailto:A.Castro@thphys.uni-heidelberg.de}{A.Castro@thphys.uni-heidelberg.de}} \qquad \textbf{Adrian Tanasa}$^b$\footnote{Email: \href{mailto:ntanasa@u-bordeaux.fr}{ntanasa@u-bordeaux.fr}}\\[3mm]
{\small $^a$  Institute for Theoretical Physics, Heidelberg University, Heidelberg, Germany.}\\
{\small $^b$ LaBRI, Université de Bordeaux, Talence, France.}}
\date{ }

\begin{document}

\maketitle
\begin{abstract}
\noindent
We implement numerical techniques to simulate $D$-random feuilletages,
candidates for higher-dimensional 
random geometries 
introduced in
L. Lionni and J.-F. Marckert, {\it Math. Phys. Anal. Geom.} {\bf 24} (2021) 39.
Using finite-size scaling techniques,
our approach allows to 
give a numerical estimation of
the Hausdorff dimension $d_H$ of these feuilletages. The results obtained are compatible with the formal result known for the Brownian map, which corresponds to the $D=2$ random feuilletage. For the $D=3$ case, our numerical study finds a good agreement with the conjectured value $d_H=8$.
\end{abstract}

\section{Introduction}

The construction of random geometries in dimension higher than two has been a very active line of research in the last years
\cite{Sturm2022,Schiavo2025,Ding:2023ufn}.
Although two-dimensional models are now well understood due to powerful combinatorial and probabilistic techniques, but also due to connections to quantum field theory, higher-dimensional analogs remain elusive. 
In particular, one would like to identify natural candidates for scaling limits of random discrete structures in dimension $D \geq 3$, which would be expected to play a role analogous to the role played by the Brownian sphere in $D=2$.

In dimension two, scale invariance is enhanced to conformal invariance, and thus random geometry is known to be tied to conformal invariant structures such as the Gaussian free field and Schramm-Loewner evolution. 
This has enabled the rigorous construction of Liouville Quantum Gravity (LQG) \cite{Duplantier:2008prc,duplantier2014liouville}
and its identification as the scaling limit of uniform random planar maps. 
% It is a rare case where probabilistic, combinatorial, and QFT perspectives coincide, 2D is special

On the other hand, 
a key breakthrough in the study of planar maps came from the discovery of bijections between rooted quadrangulations and labeled trees
\cite{cori1981planar,Schaeffer1998ConjugaisonDE,bouttier2004}.
These bijections provide combinatorial encodings that are particularly useful for the enumeration of planar maps and the study of their scaling limits. 
In particular, these bijections opened the way for the construction of the Brownian map as the universal limit of some families of random planar maps.

With the aim of looking for analogous scaling limit objects in higher dimensions, recent work has focused on extending tree-like encodings and bijective constructions to richer classes of discrete objects \cite{Lionni:2019bzb, Budd:2022dhq,Budd:2022qlr,Budd:2025cvn}. 
A particularly promising proposal is the one of $D$-random \emph{feuilletages}, $\mathbf{R}_n[D]$, introduced in \cite{Lionni:2019bzb}, which uses the encoding of planar maps by iterated trees. 
The scaling limit of $\mathbf{R}_n[D]$, denoted $\mathbf{r}[D]$, corresponds to the continuum random tree \cite{aldous} (for $D=1$), the Brownian map \cite{legall2012scaling,miermont2013brownian,marckert2006limit} (for $D=2$), and conjectural higher-dimensional analogues for $D>2$. 

Within this $D$-random feuilletages proposal, we focus in this paper on the first higher-dimensional case beyond the Brownian map, namely the $D=3$ case.
We implement a numerical simulation of random discrete feuilletages $\mathbf{R}_n[3]$ for increasing volumes $n$ and study the large-scale behavior of their distance profiles.
We adapt the techniques of \cite{Barkley:2019kvp} and \cite{Budd:2022dhq} to obtain a numerical estimation of the Hausdorff dimension 
$d_H$
by finding a consistent scaling of distance histograms as functions of the graph size $n$ (the number of vertices of the graph).
Our results provide the first numerical evidence for the conjectured value 
\[
   d_H(\mathbf{r}[3]) = 2^3 = 8 ,
\]
and illustrate the feasibility of numerical investigations of higher-dimensional random feuilletages. 

The strategy we adopt in this paper is to first determine the Hausdorff dimension $d_H$ for the trees $\mathbf{T}^D$, for $D=2$ and $D=3$ as well as for uniform quarangulations (corresponding to $\mathbf{R}[2]$), in order to verify the numerical consistency of the simulations against the exact theoretical predictions knwon for $D=2$.% $d_H(\mathbf{T}^D) = 2^D$.
More broadly, our work illustrates that the framework of iterated foldings can be probed using numerical techniques, thus opening the way to a systematic study of the geometry and scaling properties of $\mathbf{R}[D]$ for $D > 3$.

This methodology is not only of mathematical interest {\it per se}, but it also provides a tool to investigate higher-dimensional random geometries that may play a role in fundamental physics. In particular a central challenge in theoretical physics is to understand spacetime at high-energy regimes such as near the Big Bang or black hole singularities, where both classical general relativity and the Standard Model fail to make predictions. A perturbative quantization of the metric field leads to uncontrollable fluctuations at the Planck scale, showing that quantum gravity requires genuinely non-perturbative, background-independent approaches. One such strategy is to approximate the path integral over spacetimes as a sum over discrete geometries built from elementary building blocks, with the hope that a well-defined continuum limit emerges as their size goes to zero. Evidence from quantum field theory approaches and dynamical triangulations suggests that such limits may exist, but the geometry of the resulting continuum is still poorly understood beyond two dimensions \cite{Ambjorn:1991wq,Boulatov:1991hg,Hagura:1995zf,Egawa:1997fg}. In particular, identifying whether higher-dimensional random geometries behave like smooth manifolds, fractals, or new universality classes is crucial to assess their suitability as models of quantum spacetime.

Motivated by the success of Liouville quantum gravity in two dimensions, it is natural to expect that suitable ensembles of random geometries in three and four dimensions will play a comparable role in a consistent quantum theory of gravity. Given the recent proposal of \cite{Lionni:2019bzb} of iterated folding constructions as natural higher-dimensional analogues of $\gamma = \sqrt{8/3}$-LQG (a.k.a. the Brownian map), we investigate the three-dimensional case by simulating such random geometries and by giving a numerical estimation of their Hausdorff dimension. This geometric observable plays a central role: it characterizes the effective large-scale geometry of the model and serves as a critical exponent controlling the scaling of volumes with distance or, equivalently, the two-point function with geodesic distance \cite{Ambjorn:1995dg}. Our results thus aim to assess whether feuilletages provide a viable framework for extending the successes of two-dimensional quantum gravity to higher dimensions.

\medskip

The rest of this paper is organized as follows. In Section~\ref{sec:feuilletages} we review the construction of random discrete feuilletages $\mathbf{R}_n[D]$ and their known scaling limits. 
Section~\ref{sec:simulations} describes our simulation setup, including the algorithm that generates random feuilletages.
We also explain here the numerical approach we use.
%to estimate 
%and how we measure distances, as well as the 
%Hausdorff dimension for the feuilletages. 
Our numerical results are presented in Section~\ref{sec:results}, where we analyze distance-scaling in order to numerically estimate the Hausdorff dimension of $\mathbf{r}[3]$. Finally, in Section \ref{section:conclusions}, we comment on the outlook of our results and possible extensions.

\section{Review of the Iterating Folding of $D$-random Feuilletages}
%: Iterated Foldings and Higher-Dimensional Random Geometry}
\label{sec:feuilletages}

We first recall some basic definitions as well as the construction of trees and maps in order to motivate their higher dimensional analogs.

\medskip

A graph $\Gamma_n$ is a pair $(V, E)$ where $V$ is the set of $n$ vertices and $E$ is the set of edges. 
The geodesic graph distance between two vertices is the number of edges of the shortest path between them. 

A planar map can be seen as an equivalence class\footnote{Modulo orientation preserving homeomorphisms of the sphere.} of connected graphs embedded in a sphere such that the embedded edges do not cross and all regions of the surface bounded by edges are topological disks. These disks are called faces of the map. The degree of a face is given by the number of edges that bound it. A map is said to be rooted if it has a distinguished oriented edge. 
%The easiest example of a map is the map with one face which is called a tree.

\subsection{Trees}
\begin{figure}[htbp]
\centering
\includegraphics[width=\textwidth]{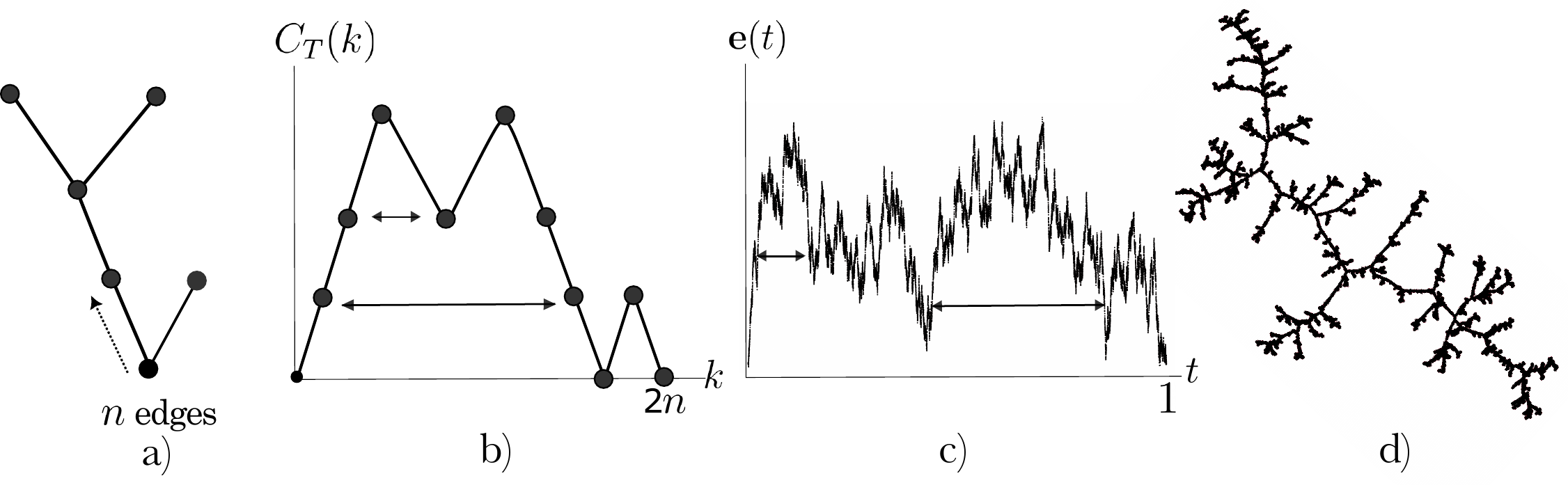}
\caption{a) Rooted plane tree with $n$ edges, the root vertex is at the bottom. b) Contour function obtained by recording the distance of each vertex from the root vertex. c) Brownian excursion obtained in the scaling limit. An example of the distance identification \eqref{eq:identif_brow_exc} is shown in horizontal arrows. d) Illustration of the Continuuos Random Tree. \label{fig:contourtocrt}}
\end{figure}

Consider a rooted plane tree with $n$ edges, $T_n$, together with its graph distance. The \textit{contour sequence} $C_T$ of this tree is obtained by recording the distance of each vertex from the root starting from the root to its leftmost neighbor and ending from its rightmost neighbor to the root. $C_T(k):[0,2n]\rightarrow\mathbb{R}_{\geq 0}$ (See Figure~\ref{fig:contourtocrt}b). This leads to a positive one-dimensional discrete walk s. t. $C_T(0)=C_T(2n)=0$. If we consider a rooted plane tree sampled uniformly at random, $C_T$ has the law of a Dyck path of length $2n$. Moreover, this is a bijection. This means that given a contour sequence, one can construct the rooted plane tree  that has such a Dyck path as a contour sequence. 
Furthermore, this bijection facilitates the enumeration of rooted plane trees, giving as a result the well-known Catalan numbers. 
This allows to study the asymptotic enumeration of these trees by looking at the large $n$ asymptotics of the Catalan numbers. The result is 
\begin{equation}
    Z^{trees}_n\stackrel{n\to\infty}{\sim} \frac{1}{\sqrt{\pi}}\;4^n n^{-3/2}.\label{eq:largen_tree}
\end{equation}
This power-law behavior is characterized by the exponent $\gamma_s-2=-3/2$, where $\gamma_s$ is called the \textit{string susceptibility}, in this case $\gamma_s=-1/2$. It characterizes the universality class of a large class of trees. 
%The reason is that a generic (appropriately normalized) random walk converges to a Brownian excursion $\mathbf{e}(t)$ \cite{revuz2004continuous} (see Figure~\ref{fig:contourtocrt}c). 
%The scaling of the contour function and its argument in \eqref{eq:to_browexcur} as $n\rightarrow\infty$ means that we are taking the number of steps to infinity while decreasing the size of the steps to be infinitesimal.     
\\
The bijection between discrete trees and Dyck paths can be lifted to their continuum setting. This allows to map the Brownian excursion, $\mathbf{e}$, back to a tree by identifying points at distance zero given the distance function
\begin{equation}
    d_{\mathbf{e}}(s,t)=\mathbf{e}(s)+\mathbf{e}(t)-2\inf_{u\in[s,t]} \mathbf{e}(u)\qquad 0\leq s,t \leq 1,\label{eq:identif_brow_exc}
\end{equation}
The last term restricts the identification when the excursion drops below $\mathbf{e}([s,t])$ (see horizontal arrows in Figure \ref{fig:contourtocrt}c). This object is called the Continuous Random Tree (CRT) \cite{aldous}. 

%In general, any such excursion $X : [0,1] \to \mathbb{R}_{\geq 0}$ naturally gives rise to a continuous metric space: the \emph{real tree} given by the unit interval $[0,1]$ with metric 
%\begin{equation}
 %   d(s,t) = X(s) + X(t) - 2 \inf_{u\in [s,t]}X(u),
%\end{equation}
%where it is understood that we identify $s$ and $t$ whenever $d(s,t)=0$.

\subsection{Maps} 
Going one step further in `dimensionality', we briefly review the case of maps and the celebrated  Cori-Vaqueli-Schaeffer (CVS) bijection \cite{cori1981planar,Schaeffer1998ConjugaisonDE} which allows us to relate planar maps out to (well) labelled trees.

\begin{figure}[H]
\centering
\includegraphics[width=0.6\textwidth]{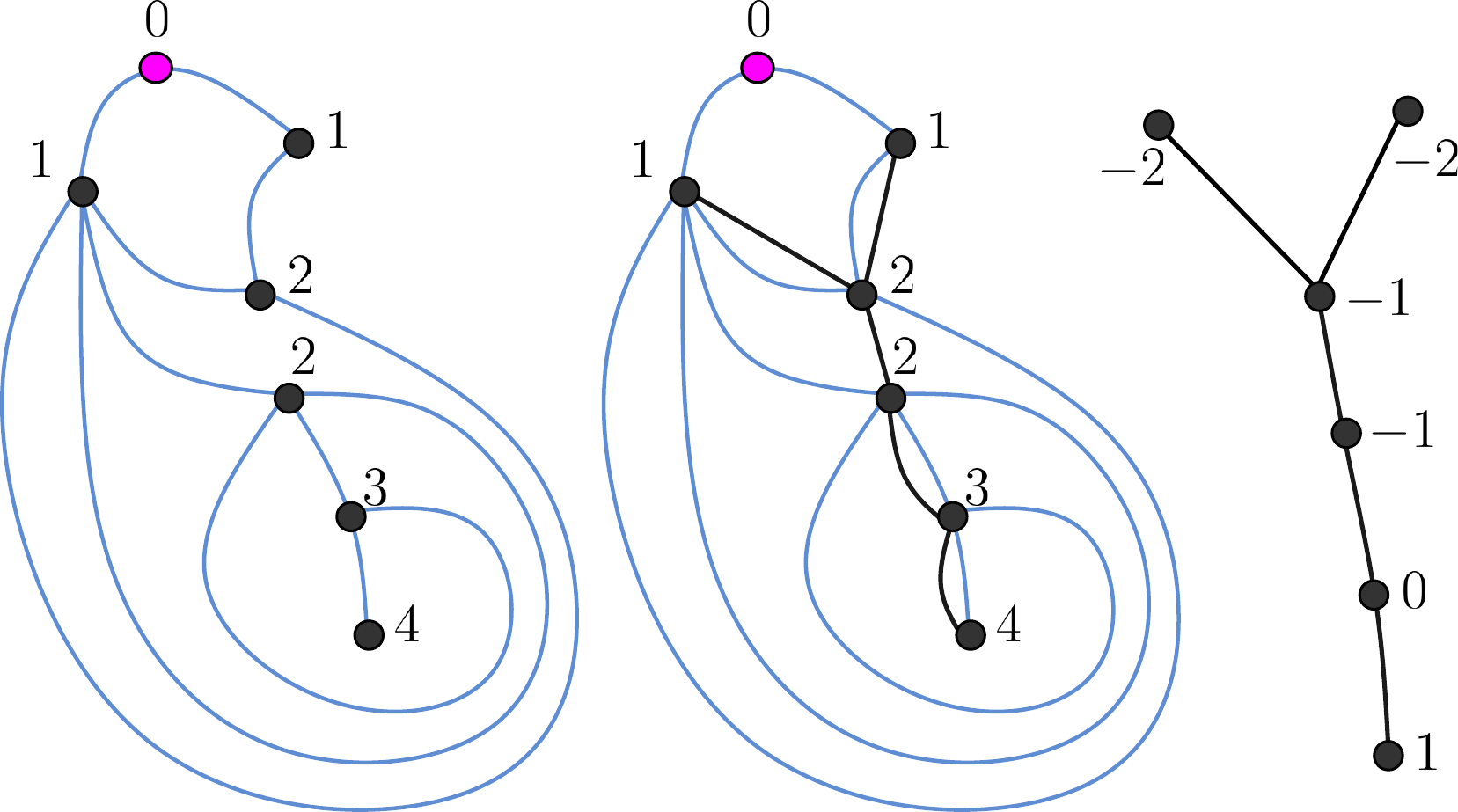}
     \put(-230,-10){a)}
     \put(-110,-10){b)}
     \put(-15,-10){c)}
\caption{a) Pointed planar quadrangulation with 6 faces. The root vertex is shown in fuchsia. Each vertex is labeled by its graph distance to the root. b) The spanning tree (in black) is obtained by the CVS bijection with positive labels. c) Labeled plane tree with 5 edges and vertex labels such that between adjacent vertices the labels differ by $-1$, $0$ or $+1$.\label{fig:CVS}}
\end{figure}

Let $M_n$ be a pointed planar quadrangulation with $n$ faces. This is a planar map with a marked vertex (called root vertex) and where every face has degree 4. Assign to each vertex a label given by the graph distance from the root vertex (Fig. \ref{fig:CVS}a). The labeled quadrangulation has two types of faces: those with vertices with labels $(k, k+1, k+2, k+1)$ and those with labels $(k, k+1, k, k+1)$. Draw an edge inside of every face of the map using the following rule: for faces $(k, k+1, k+2, k+1)$ the new edge joins vertices $k+1$ and $k+2$; for the faces $(k, k+1, k, k+1)$, the new edge joins the two vertices $k+1$. This process produces a tree with $n$ edges joining the vertices of the map $M_n$ minus the root vertex labeled with positive integers (Fig. \ref{fig:CVS}b). Now, in order not to restrict the labels to be positive, we can always take $k_{max}$ to be the maximum label of this tree, subtracting $k_{max}-1$ from all the labels of the resulting tree, we obtain a \textit{labeled} plane tree with $n-1$ edges (Fig. \ref{fig:CVS}c). That is a plane tree where each vertex is labeled by an integer number and such that the labels between adjacent vertices differ by $-1$, $0$ or $+1$. This defines a bijection\footnote{Up to a factor of $2$ accounting for the orientation of the root edge and a factor of $n+2$ accounting the choice of root vertex.} between rooted quadrangulations with $n$ faces and rooted labeled plane trees with with labels varying at most 1 along its edges.\\
\noindent This is known as the CVS bijection and allows one to enumerate quadrangulations with $n$ faces by enumerating rooted plane trees plus the number of allowed labelings. This leads to the well-known result on the asymptotic enumeration of quadrangulations
\begin{equation}
    Z^{quad}_n\stackrel{n\to\infty}{\sim} \frac{2}{\sqrt{\pi}}\;12^n n^{-5/2},\label{eq:largen_map}
\end{equation}
where %it is straightforward to see that, in this case, 
$\gamma_s=1/2$.\\
Given this bijection between labeled trees and planar maps, the scaling limit of the latter can be studied. As mentioned above, the scaling limit of the tree is well-defined and is the celebrated CRT. Its labeling corresponds to another discrete walk on the plane and its scaling limit is a Brownian snake. Using the `continuum' analog of the CVS bijection in these objects the Brownian map was constructed \cite{LeGall:2006aea}. The Brownian map \cite{legall2012scaling,miermont2013brownian,marckert2006limit} is the universal scaling limit of large random planar maps \cite{LeGall:2006aea,LeGall2011UniquenessAU}.
It arises as a random metric space $(\mathbf{r}[2],d)$ of Hausdorff dimension~$4$, homeomorphic to the sphere~$\mathbb{S}^2$, and is encoded via the CRT decorated with spatial labels given by a Brownian snake.\\

\medskip

A natural question to ask is whether analogous universal limits exist in higher dimension and, moreover, if these can be constructed in a similar way. It is in this spirit that in \cite{Lionni:2019bzb} the $D$th random feuilletage was introduced.

\subsection{The arbitrary $D$ case}

\begin{figure}[h]
\centering
\includegraphics[width=\textwidth]{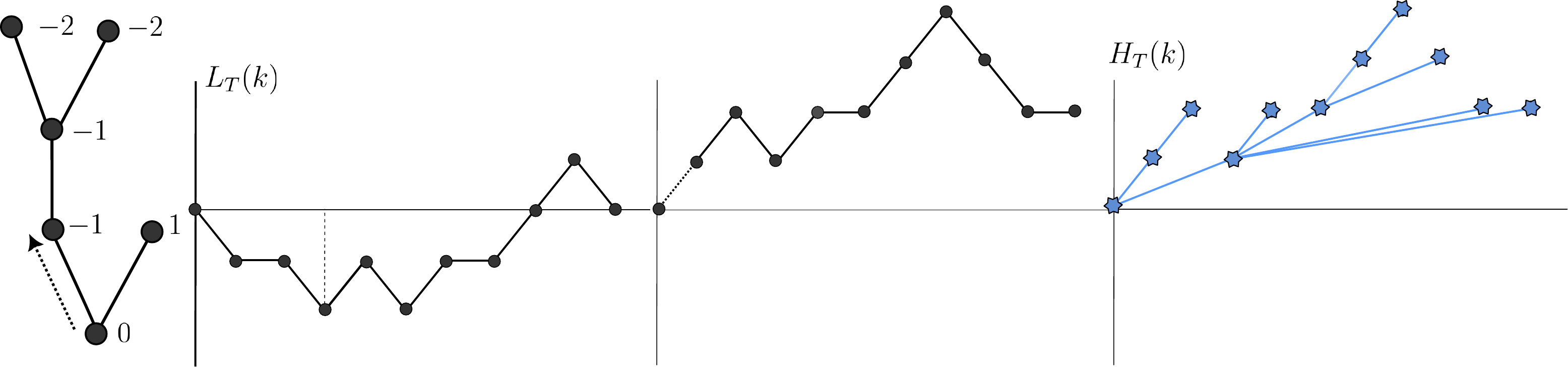}
     \put(-406,-10){a)}
     \put(-325,-10){b)}
     \put(-200,-10){c)}
     \put(-70,-10){d)}
\caption{a) Uniform labeled plane tree with $5$ edges $\mathbf{T}_5^1$. b) Label process of $\mathbf{T}_5^1$. c) Conjugation of the label process. d) Height process of the random plane tree with $10$ edges, $\mathbf{T}_{10}^2$. \label{fig:CtoH}}
\end{figure}
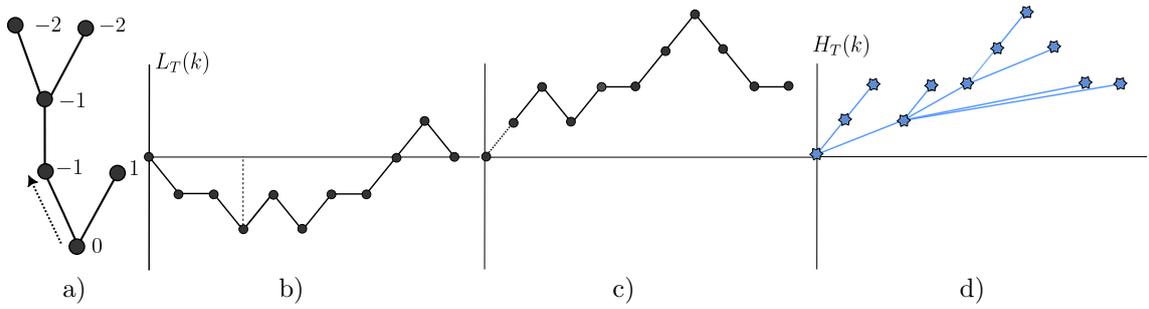

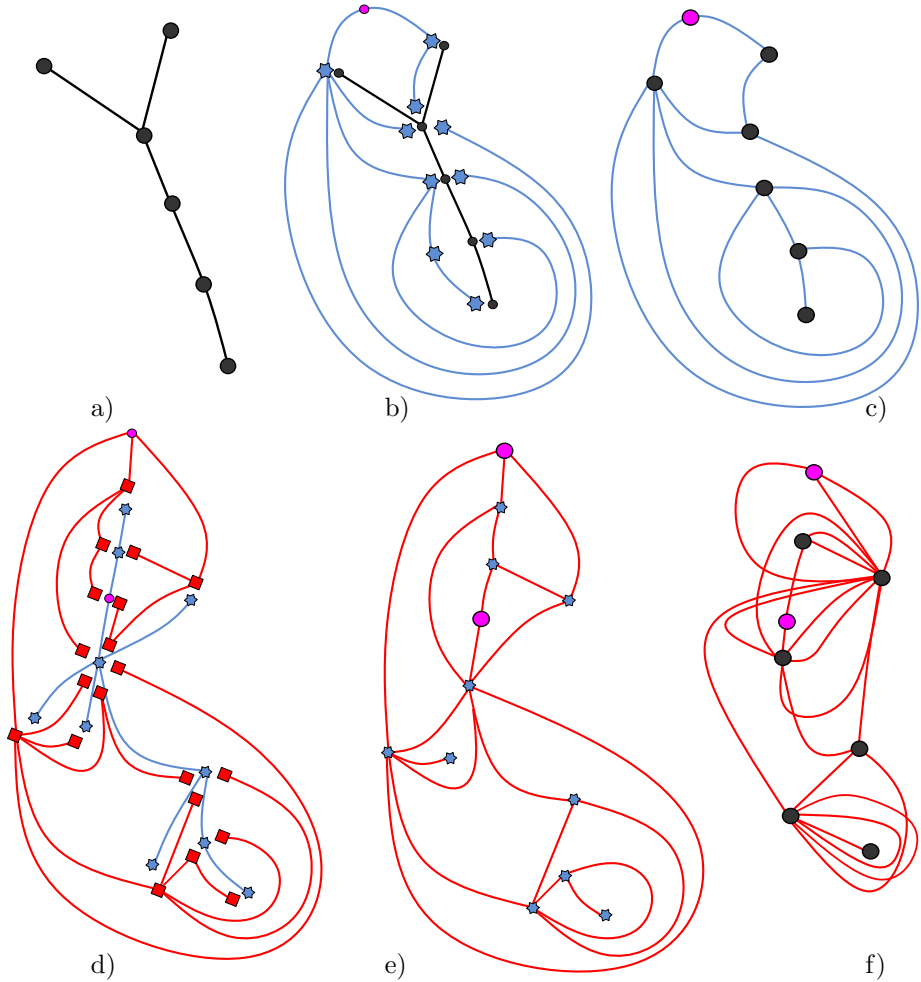
\begin{figure}[H]
\centering
\includegraphics[width=0.8\textwidth]{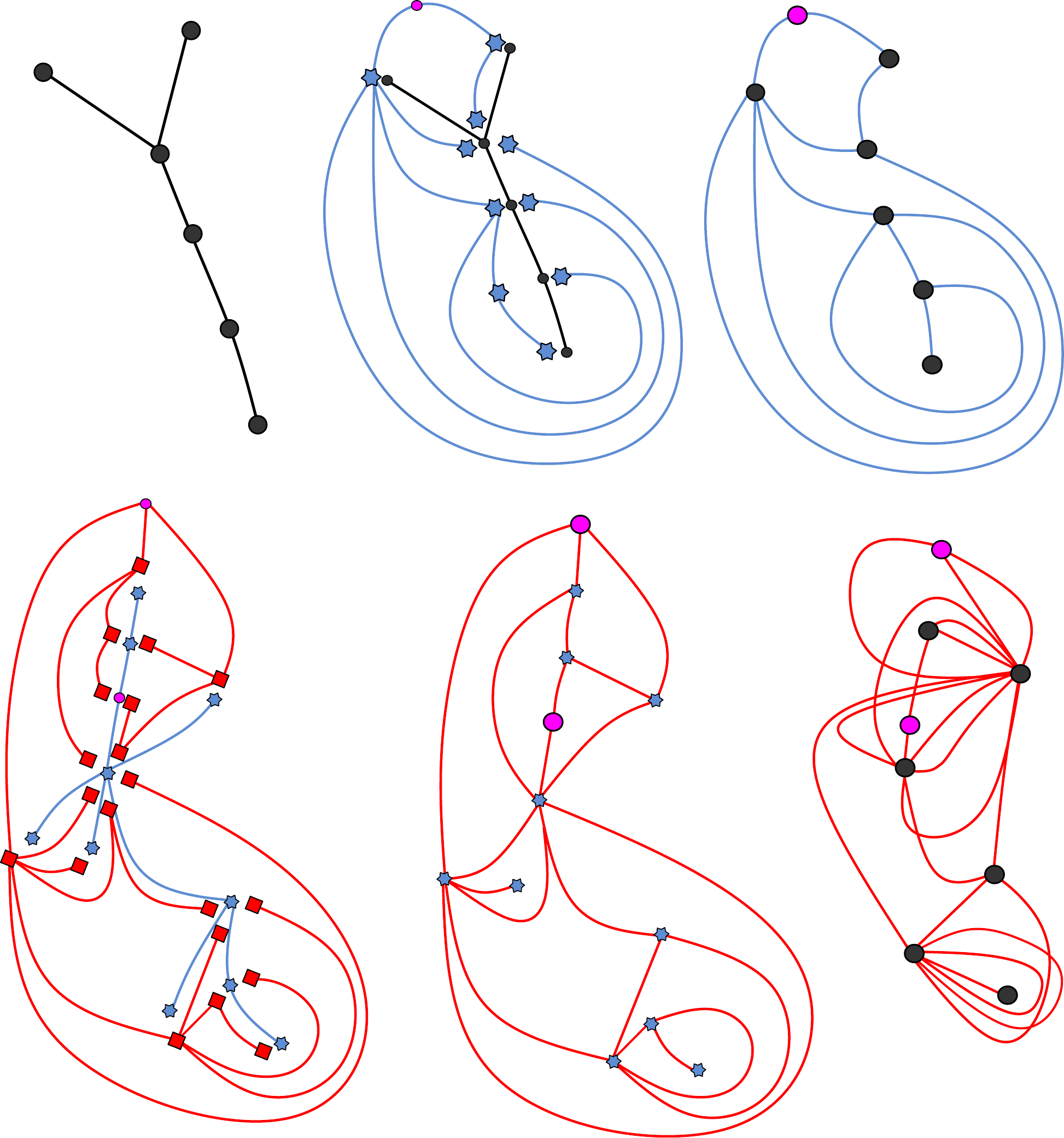}
     \put(-310,210){a)}
     \put(-200,210){b)}
     \put(-20,210){c)}
     \put(-310,0){d)}
     \put(-200,0){e)}
     \put(-20,0){f)}
\caption{a) $(D=1)$-random discrete feuilletage $\mathbf{T}^1$. b) Identification of the vertices of the random plane tree $\mathbf{T}^2$ induced by the corners of $\mathbf{T}_1$. c) Corresponding uniform pointed planar quadrangulation or $(D=2)$-random discrete feuilletage. d) Identification of the vertices of the random plane tree $\mathbf{T}^3$ induced by the corners of $\mathbf{T}^2$. e) Corresponding random planar map with two-marked vertices. f) $(D=3)$-random discrete feuilletage obtained from  identifying vertices of the random plane trees $\mathbf{T}^3$ (according to the corners of $\mathbf{T}^2$) and $\mathbf{T}^2$ (according to the corners of $\mathbf{T}_1$).\label{fig:feu}}
\end{figure}

%The construction of these feuilletage starts with 
The $D$th random discrete snake of size $n$ is given by
\[
\mathbf{BS}_n[D]=\left([\mathbf{C}_n^{(1)},\mathbf{L}_n^{(1)}],\dots, [\mathbf{C}_n^{(D)},\mathbf{L}_n^{(D)}] \right),
\]
where $\mathbf{C}_n^{(1)}$ is a uniform Dyck path of $2n$ steps which corresponds to the contour process of a uniform planar tree $\mathbf{T}^{1}_n$ with $n$ edges (Fig. \ref{fig:contourtocrt}a)). 
Moreover, $\mathbf{L}_n^{(1)}$ is a random label process  induced by a standard branching random walk on the tree $\mathbf{T}^1_n$ (Fig. \ref{fig:CtoH}b)).  For $j>1$, the pair ($\mathbf{C}_n^{(j)},\mathbf{L}_n^{(j)}$) corresponds to the contour and labeling induced by a standard branching random walk of a random planar tree $\mathbf{T}^{j}_n$ with $2^{j-1} n$ edges (see, for example, Fig. \ref{fig:CtoH}).\\
The $D$th \emph{random discrete feuilletage}, denoted 
$\mathbf{R}_n[D]$, is
a random metric space with $n+D$ nodes obtained by iterated folding of discrete trees as follows: For all $j\geq 2$, the nodes ($\mathbf{a}_n^{(j-1)}+c-1$ mod $2^{j-1} n$) and ($\mathbf{a}_n^{(j-1)}+c'-1$ mod $2^{j-1} n$) of the tree $\mathbf{T}^{(j)}_n$ are \textit{identified} if the corners $(c,c')\in [1,2^{j-1} n]^2$ are corners of the same node in $\mathbf{T}^{(j-1)}_n$ (see Fig. \ref{fig:feu}).

Let us remind here a couple of particularly important properties of these random discrete feuilletage $\mathbf{R}_n[D]$: 
\begin{itemize}
    \item Its edges coincide with those of the tree $\mathbf{T}_n^{(D)}$.
    \item Its vertices correspond to those of $\mathbf{T}_n^{(1)}$ plus the root vertices of the trees $\mathbf{T}_n^{(j)}$ with $2\leq j\leq D$.
\end{itemize}

\begin{table}[H]
\centering
\begin{tabular}{|c|c|c|}
\hline
 \textbf{Random discrete object} & in bijection with & \textbf{Scaling limit} \\
 \hline
 Plane tree  & Random contour sequence & Continuum Random Tree (CRT) \\
 Quadrangulation & Random labelled tree & Brownian map\\
 $D$th discrete feuilletage  & $D$th discrete Brownian snake & $D$th random feuilletage\\
\hline
\end{tabular}
\caption{Relations between trees, quadrangulations and the $D$-th random feuilletage.}
\label{fig:schematic_D_feuill}
\end{table}
\noindent The asymptotic enumeration of rooted $D$th discrete feuilletages was shown (see again \cite{Lionni:2019bzb}) to be 
\begin{equation}
    Z_n^{(D)}\sim c_D \lambda_D^n\; n^{\gamma_s[D]-2}.
\end{equation}
The string susceptibility of the $D$th discrete feuilletage is now 
\begin{equation}
    \gamma_s[D]=\frac{3}{2}-D.
\end{equation}
This reduces to the well-known values of $\gamma_s[1]=-1/2$ for random trees \eqref{eq:largen_tree} (branched polymers) and $\gamma_s[2]=+1/2$ for uniform quadrangulations \eqref{eq:largen_map} (pure gravity). \\
As $n \to \infty$, one has the convergence\footnote{This convergence holds in a functional sense. This convergence is weaker than the Gromov-Hausdorff convergence for the Brownian map, but it provides a rigorous approximation scheme.} of the $D$th discrete feuilletage\footnote{More precisely, of its normalized and pointed version. However, this does not affect the distance statistics we study in this work.} $\mathbf{R}_n[D]$ to the $D$th random feuilletage $\mathbf{r}[D]$. 
The metric spaces 
\[
   \mathbf{r}[D], \qquad D \geq 0,
\]
correspond to
\begin{itemize}
    \item $\mathbf{r}[0]$: the unit circle,
    \item $\mathbf{r}[1]$: CRT,
    \item $\mathbf{r}[2]$: the Brownian map,
    \item $\mathbf{r}[D]$, $D \geq 3$: new higher-dimensional random geometries. 
\end{itemize}

\noindent By construction, the upper bound for the diameter of the discrete feuilletage $\mathbf{R}_n[D]$ is
\begin{equation}
       \mathrm{diam}\,\mathbf{R}_n[D] 
       \lesssim 
       n^{1/2^D}.
\end{equation}
This comes from the fact that there is always a path to go from vertex $a$ to vertex $b$ in $\mathbf{R}_n[D]$ along the tree $\mathbf{T}_n^D$. However, the identification requested to obtain the corresponding map can bring the vertices closer together, creating `shortcuts'.  This sets a lower bound for the Hausdorff dimension of the $D$th discrete feuilletage. The Hausdorff dimension accounts for the relative scaling of volumes of balls with respect to their geodesic radius. 
Based on these diameter estimates and analogy with the $D=1,2$ cases, it was conjectured \cite{Lionni:2019bzb} that the Hausdorff dimension of $\mathbf{r}[D]$ satisfies
\begin{equation}
    \dH (\mathbf{r}[D]) = 2^D.
\end{equation}   
This is consistent with known results: $\dH (\mathbf{r}[1])=2$ for the CRT \cite{aldous} and $\dH (\mathbf{r}[2])=4$ for the Brownian map \cite{LeGall:2006aea}. 
For $D \geq 3$, this remains an open question. It is to this open question that we bring numerical evidence in the rest of this paper.

\section{Methods and Numerical Implementation}\label{sec:simulations}

In this section, we introduce the %numerically-motivated 
Hausdorff dimension numerical estimator we use (based on \cite{Barkley:2019kvp} and \cite{Budd:2022qlr}) as well as a description of the numerical implementation. 
%in the code \cite{code}.

\subsection{Hausdorff dimension estimator}

Let $\Gamma_n$ be a graph with $n$ vertices equipped with the canonical graph distance $\hat{G}_n$. If the metric space given by the pair $(\Gamma_n, \hat{G}_n)$ has a scaling limit, there exists a positive real number $d_H$  such that  the limit 
\begin{equation}
    \left(\Gamma_n, n^{-1/d_H} \hat{G}_n\right)\xrightarrow[]{n\rightarrow \infty} \left(\Gamma, \hat{G}\right)
\end{equation}
exists in the Gromov–Hausdorff sense.

In order to find this exponent numerically, we simulate $q$ graphs of $n$ vertices and measure graph distances in the following way. Let $x_0, x_1$ uniformly random vertices of the graph $\Gamma_n$ and let $\hat{G}_n(x_0,x_1)= r_n$. Then, the Hausdorff dimension $d_H$ is the exponent such that the limit 
\begin{equation}
    n^{-1/d_H} r_n \xrightarrow[]{n\rightarrow \infty} r
\end{equation}
exists.

In order to make the numerical estimation statistically relevant, we need to consider $q\gg 1$, {\it i.e.} we need to consider a significantly large number of samples, and we also need to take $n\gg 1$ to approximate the scaling limit. This information can be efficiently encoded in a normalized histogram $\rho_n(x)=\mathbb{P}(r_n=x)$ for $x\in \mathbb{R}_{\geq 0}$. For a fixed value $x$, $\rho_n(x)$ is the probability that two vertices of $\Gamma_n$ are at distance $x$.
In this setting, we assume that the existence of $d_H$ implies the existence of the scaling limit of $n^{1/d_H} \rho_n (n^{1/d_H} x)$. To justify this assumption, we first check its validity 
for the well-known $D=2$ (uniform quadrangulations) case.

Finally, given that the formal limit $n\rightarrow\infty$ is unattainable in numerical simulations, we use a reference volume $n_0\gg 1$. Then, the Hausdorff dimension exists if there exists parameters $k_n>1$ for $n_0>n$ and $k_{n_0}=1$ such that
\begin{equation}
    k_n^{-1} \rho_n \left(k_n^{-1} x\right)\hspace{1cm} \mathrm{fits} \hspace{1cm} \rho_{n_0}(x).\label{eq:fit_hist}
\end{equation}
More precisely, for each $n<n_0$ we determine the fit parameters $k_n$ and $s_n$ that minimize the integrated square deviation between $k_n^{-1}\rho_n(k_n^{-1}(x+s_n)-s_n)$ and $\rho_{n_0}$. 
The shift $s_n$ is included to compensate discretization effects. 
In order to take into account the dependence of $s_n$ on $n$, we fit $\rho_n$ twice: first time to find the values of $s_n$ and to obtain the mean $s$, and the second time we fit  $k_n^{-1}\rho_n(k_n^{-1}(x+s)-s)$ to obtain the values of $k_n$.
The relation between $k_n$ and $n$ implies that in the limit $n\rightarrow \infty$, the parameter $k_n$ behaves as 
\begin{equation}
    k_n \sim (n/n_0)^{-1/d_H}.\label{eq:d_Hsim}
\end{equation}
% Therefore, 
% \begin{equation}
%     \log( k_n) \sim  -\frac{1}{d_H} \log (n/n_0).\label{eq:d_Hsim}
% \end{equation}
%Following this analysis which proved efficient in \cite{Barkley:2019kvp} and \cite{Budd:2022qlr}, we find an 
Thus, 
a numerical 
estimate of the Hausdorff dimension is obtained by fitting this curve. To be more precise, one can parameterize the deviations from \eqref{eq:d_Hsim} by
\begin{equation}
    k_n=\left(\frac{n}{n_0}\right)^{-1/d_H}\left(a+b\left(\frac{n}{n_0}\right)^{-\delta}\right)\label{eq:num_fit_dh}
\end{equation}
where $a\sim 1$, $|b|\ll 1$ and $\delta$ of order $1/d_H$.

Additionally, in order to corroborate the robustness of our results, we use different deciles as introduced in \cite{Fredes2021ModelsOR}. That is, to perform the fit \eqref{eq:fit_hist}, we do not use the whole histogram but $75\%$, $50\%$ and $25\%$ from the maximum value of the histogram. 

\subsection{Numerical implementation}

In this subsection we present the  \texttt{C++} algorithm used to 
%simulate the random feuilletages $\mathbf{R}_n[D]$ and to 
estimate the %large-scale geometric properties, in particular 
the Hausdorff dimension of the feuilletages. The implementation is based on successive foldings of Dyck paths and their associated label processes.

The starting point of each simulation is a Dyck path of length $2n$, generated uniformly at random. In the code this is performed by the function 
\texttt{SimulationDyck}, which outputs an integer array $V[0]$ of length \texttt{size[0]}. This corresponds to the contour function of the first rooted plane tree.
To each tree we assign a label process $\mathrm{Eti}$, constructed iteratively via the function \texttt{ProcessusEtiquette}. 
The pair $(V[i],\mathrm{Eti}[i])$ forms the basic building block at iteration $i$.

The core of the algorithm is the iterative folding procedure implemented in the function \texttt{IteratedMap}. At each iteration $i$, the following operations are performed:
\begin{enumerate}
    \item The current label process $\mathrm{Eti}[i]$ is conjugated to a new height function $H1[i+1]$ using \texttt{Conjug}.
    \item A new contour path $V[i+1]$ is generated from $H1[i+1]$ via a function \texttt{FromHtoV}, producing the next iteration of the tree.
    \item The representative nodes $\mathrm{Rep}[i]$ obtained from the function  \texttt{NodesRepresentants}, which encodes how each vertex is identified under folding, are projected to the previous level through the recursion
    \[
        Rg[i][j] =
        \begin{cases}
            0, & j=0, \\
            Rg[i-1]\big( \, \texttt{modu}(\mathrm{Rep}[i][j] + a[i-1] - 1,\, \texttt{size}[i-1]) \, \big), & j>0,
        \end{cases}
    \]
    where $a[i]$ are the conjugacy parameters. This ensures that node identifications are consistently propagated across iterations.
    \item Intermediate arrays such as $V[i]$, $\mathrm{Eti}[i]$, and $\mathrm{Rep}[i]$ are freed to optimize memory usage.
\end{enumerate}
This procedure constructs the $\mathbf{R}[D]$ feuilletage after $(D-1)$ steps (corresponding to the parameter \texttt{ITERATION}).

Distance estimations are performed using the function \texttt{ComputeDistanceFromRandomNode}. 
At each iteration, the following operations are performed:
\begin{enumerate}
    \item 
A random vertex in the final folded map is selected as the root, and its distance is initialized to zero. 
\item 
Distances to all other nodes are iteratively updated using the tree height array $H_1$ and the representative mapping $Rg$. At each step, the distance of a node is compared to that of its neighbor along the tree. 
\item
This process produces a complete distance profile from the chosen root. The maximum distance is recorded.
\item  To gather sufficient statistics, the procedure is repeated over 10 random roots chosen independently of each other and 100000 independent map realizations using a function\linebreak \texttt{RepeatDistanceIteratedMap} from which the distance histograms are populated. This procedure was repeated to populate three (for $\mathbf{R}[3]$), four (for $\mathbf{T}_3$ and $\mathbf{R}[2]$), and resp. eight (for $\mathbf{T}_2$) statistically independent samples from which the standard deviations of our measurements where obtained.
\end{enumerate}

In each simulation, distances are measured with respect to a root node chosen at random. Concretely, a root is selected uniformly among the vertices of the initial Dyck path (before the folding procedure), i.e. from the first half of the contour representation. After folding, this choice corresponds to a specific vertex in the map, which then serves as the reference point for distance computations. The algorithm sets the distance of the root to zero and iteratively updates the distances of all other vertices. This procedure ensures that shortest–path distances in the folded map are obtained with respect to the chosen root.
Because the root is drawn uniformly from the contour prior to folding, the distribution of root choices in the final geometry is not strictly uniform across vertices. In particular, vertices with a high number of corners may be oversampled. This introduces a mild bias that can slightly shift the distance profiles and the extracted Hausdorff dimension. However, since the distance histograms show a well-behaved scaling, we regard the deviation as a systematic finite–size effect.

\section{Hausdorff dimension estimation}
\label{sec:results}

In this section, we present the numerical estimations of the Hausdorff dimensions obtained using the procedures described in Section~\ref{sec:simulations} for $D=2$ and $D=3$ random feuilletages. Our strategy is to first determine the exponent $d_H$ for the trees $\mathbf{T}^D$, in order to verify the numerical consistency of the simulations against the exact theoretical prediction $d_H(\mathbf{T}^D) = 2^D$. This preliminary step also provides a reference for estimating the numerical uncertainty on $d_H$.

For each case, we display the logarithmic fit given by \eqref{eq:num_fit_dh}. Different values of the reference volume $n_0$ are used, values chosen to balance numerical accuracy and computational feasibility: $n_0$ must be large enough for the Hausdorff dimension $d_H$ to approach its predicted asymptotic value, yet small enough for the simulations to remain tractable within reasonable cluster run times.

\newpage

\subsection{$(D=2)$-random feuilletage}

\begin{figure}[H]
\centering
\includegraphics[width=0.5\textwidth]{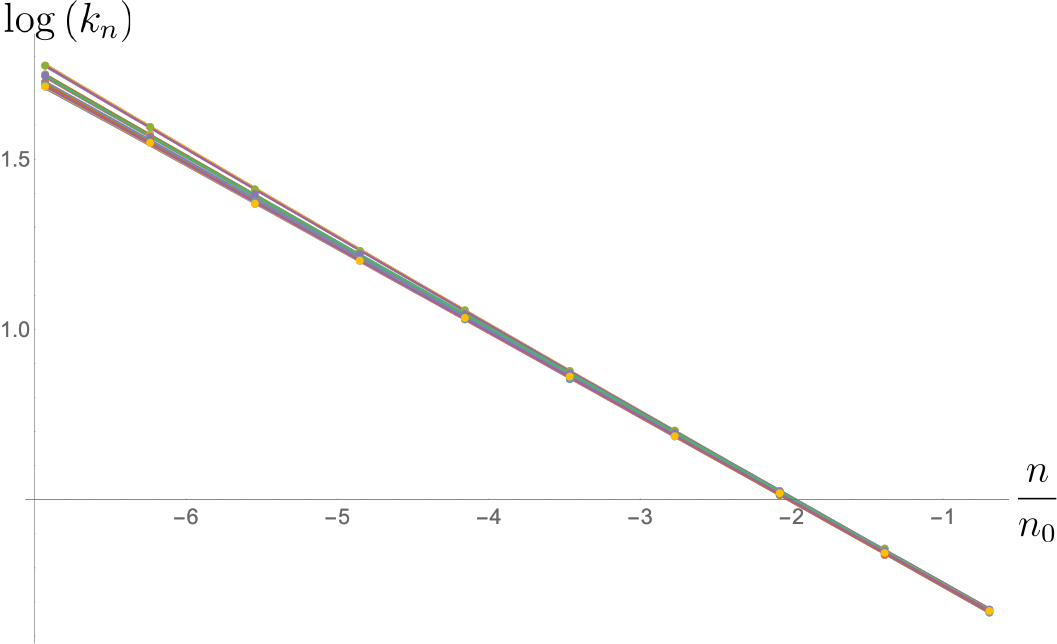}
\caption{Logarithmic plot of fit \eqref{eq:num_fit_dh}.\label{fig:fit_T2}}
\end{figure}

%As mentioned above,  we first check the consistency of our numerical results with the formal results for random labeled trees and uniform quadrangulations. 
We first study the case of random labeled trees, that is $\mathbf{T}_n^2$ with $n\in [2^{11}, 2^{21}]$.
\begin{figure}[H]
    \centering
    \includegraphics[width=\linewidth]{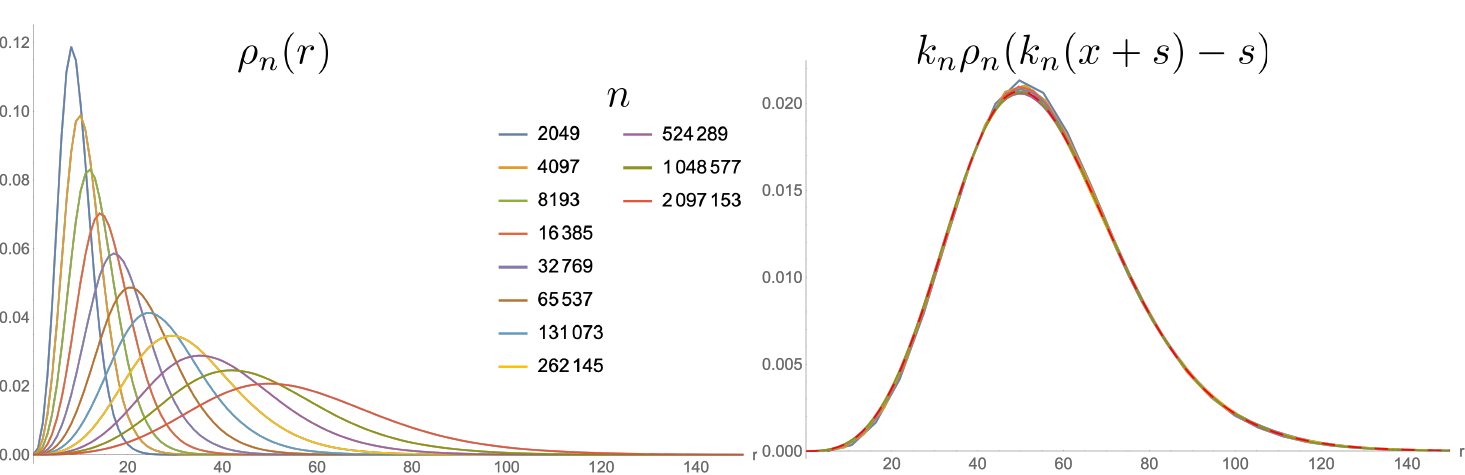}
\label{fig:hist_T2}
\caption{Left: Distance histograms $\rho_n(x)$ for $n\in [2^{11}, 2^{21}]$. Right: Example of the fit to $\rho_{n_0}$ with $n_0=2^{21}$ from which the numerical values of $k_n$ are obtained.}
\end{figure} 
The obtained values for $k_n$ in function of the size of the graph $n$ are fitted according to the ansatz \eqref{eq:num_fit_dh}. This is shown in Figure \ref{fig:fit_T2}.
%Given such fit, t
The numerical estimates for  the Hausdorff dimension $\dH$ are then  obtained. We list them in the following table, where the reference volume for each site is denoted by $n_0$, the decile corresponds to the percentage of the histogram considered for the fit from its maximum, and the error bars correspond to the standard deviation computed with respect to different batches.  
\begin{table}[h]
\centering
\begin{tabular}{|c|c|c|}
\hline
$n_0$ & Decile & \textbf{$d_H$} \\
\hline
\multirow{3}{*}{$2^{21}= 2097153$} 
   & 0.75 & $4.04489\pm 0.0509759$ \\
   & 0.50 & $4.06857\pm 0.0830516$ \\
   & 0.25 & $4.04807\pm 0.0582075$ \\
\hline
\multirow{3}{*}{$2^{20}= 1048577$} 
   & 0.75 & $4.12535\pm 0.17996$ \\
   & 0.50 & $4.08403\pm 0.177443$ \\
   & 0.25 & $4.0675\pm 0.165611$ \\
\hline
\multirow{3}{*}{$2^{19}= 524289$} 
   & 0.75 & $4.04577\pm 0.073932$ \\
   & 0.50 & $4.0427\pm 0.0681372$ \\
   & 0.25 & $4.0255\pm 0.0591166$ \\
\hline
\end{tabular}
\caption{Comparison of Hausdorff dimension estimates obtained for $\mathbf{T}^2$ with different numerical schemes given by varying the reference volume $n_0$ and the decile \cite{Fredes2021ModelsOR}.}
\label{tab:dh_tree_d2}
\end{table}

\noindent In the case of $\mathbf{R}_n[2]$ i.e. uniform quadrangulations, we considered the range $n\in [2^{11}, 2^{19}]$. We present an example of the histograms and their fits as well as the logartimic fit. 
\begin{figure}[H]
\centering
\includegraphics[width=\textwidth]{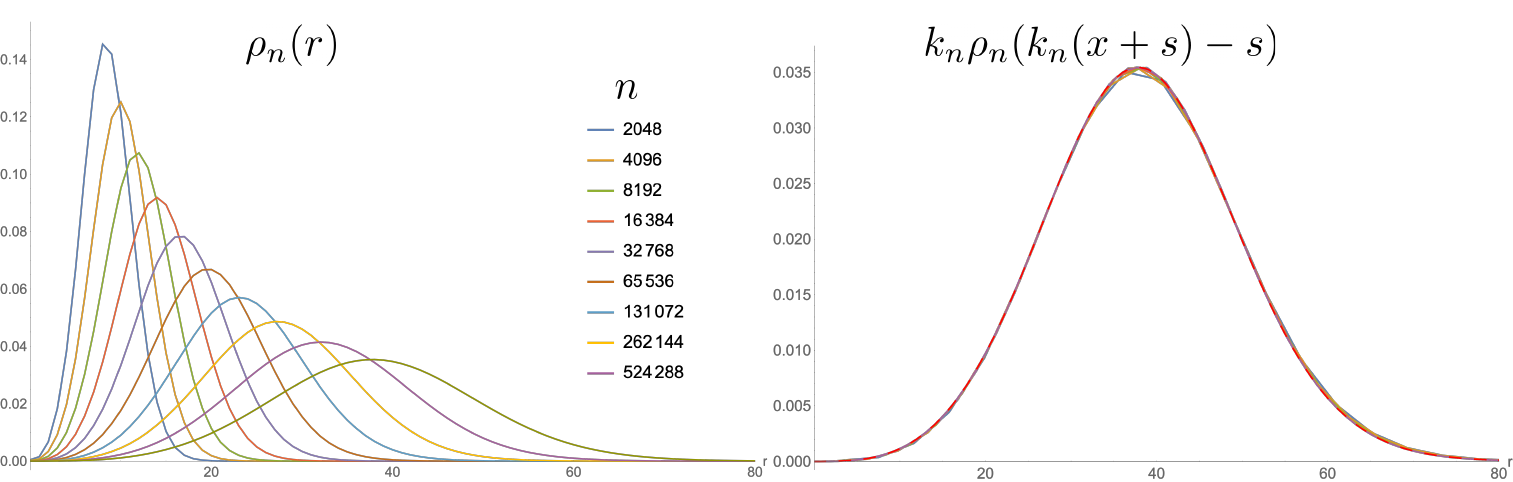}
\caption{Left: Distance histograms $\rho_n(x)$ for $n\in [2^{11}, 2^{19}]$. Right: Example of the fit to $\rho_{n_0}$ with $n_0=2^{19}$ from which the numerical values of $k_n$ are obtained.}\label{fig:hist_M2}
\end{figure}
\begin{figure}[H]
\centering
\includegraphics[width=0.5\textwidth]{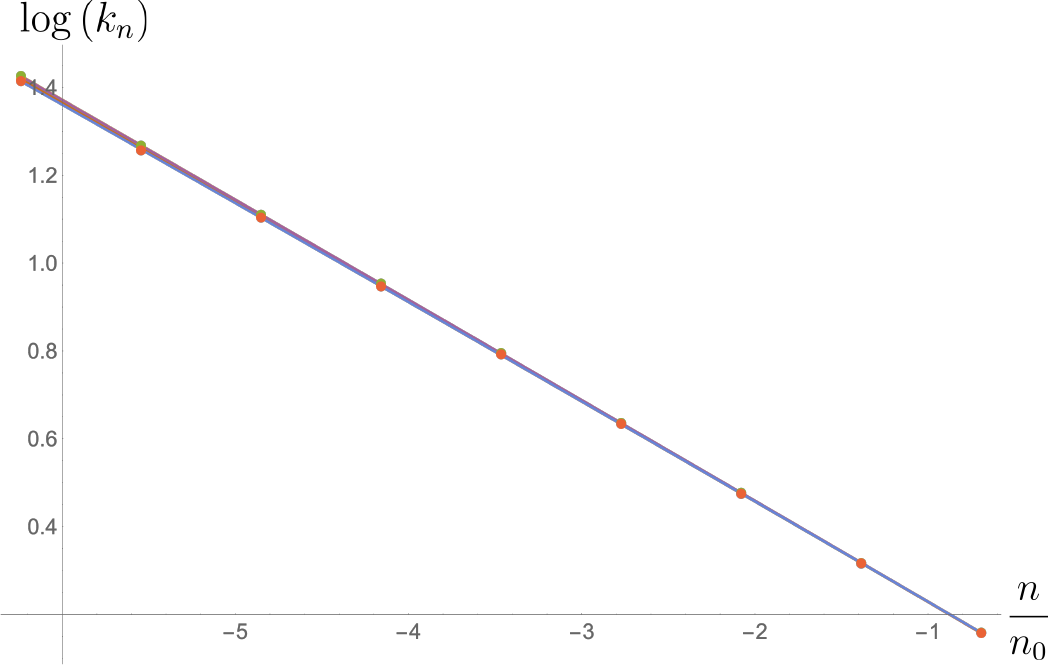}
\caption{Logaritmic plot of fit \eqref{eq:num_fit_dh}.\label{fig:fit_M2}}
\end{figure}
%Given these fits, t
The numerical estimates for the Hausdorff dimensions $\dH$ are then given in Table \ref{tab:dh_map_d2}.

\begin{table}[h]
\centering
\begin{tabular}{|c|c|c|}
\hline
Reference volume & Decile & \textbf{$d_H$} \\
\hline
\multirow{3}{*}{$2^{19}= 524289$} 
   & 0.75 & $4.09626\pm 0.0939346$ \\
   & 0.50 & $4.17282\pm 0.105188$ \\
   & 0.25 & $4.12924\pm 0.0646897$ \\
\hline
\multirow{3}{*}{$2^{18}=262144$} 
   & 0.75 & $4.157\pm 0.0866086$ \\
   & 0.50 & $4.19995\pm 0.0759246$ \\
   & 0.25 & $4.17802\pm 0.0941855$ \\
\hline
\multirow{3}{*}{$2^{17}=131072$} 
   & 0.75 & $4.35338\pm 0.0436168$ \\
   & 0.50 & $4.33901\pm 0.0452308$ \\
   & 0.25 & $4.31055\pm 0.0570731$ \\
\hline
\end{tabular}
\caption{Comparison of Hausdorff dimension estimates obtained for $\mathbf{R}[2]$ with different numerical schemes given by varying the reference volume $n_0$ and the decile \cite{Fredes2021ModelsOR}.}
\label{tab:dh_map_d2}
\end{table}

Both numerical estimates given in Tables \ref{tab:dh_tree_d2} and \ref{tab:dh_map_d2} agree with the theoretical results within the error bars as well as with previous numerical measurements \cite{Barkley:2019kvp,Budd:2022qlr}. 

Let us end this subsection by the following consideration with respect to
error tolerance. 
Note that, while the numerical estimate of $\dH$ converges very precisely to the theoretical value in the case of trees, in the case of maps, there is some finite-volume error we should be expecting in the $D=3$ case. The 'worst' numerical agreement with the theoretical value $\dH=4$ is of the order of $8.75 \%$. This is the threshold we expect to reach in the $D=3$ discrete feuilletage case. 
\newpage

\subsection{$(D=3)$-random feuilletage}

%In the case of $D=3$, w
We start in this case by simulating the tree $\mathbf{T}_n^3$ with $n\in [2^{11}, 2^{26}]$. As mentioned above, there is a formal estimate for the diameter of this object, so this serves as a testing ground 
for a consistency check.
\begin{figure}[H]
\centering
\includegraphics[width=\textwidth]{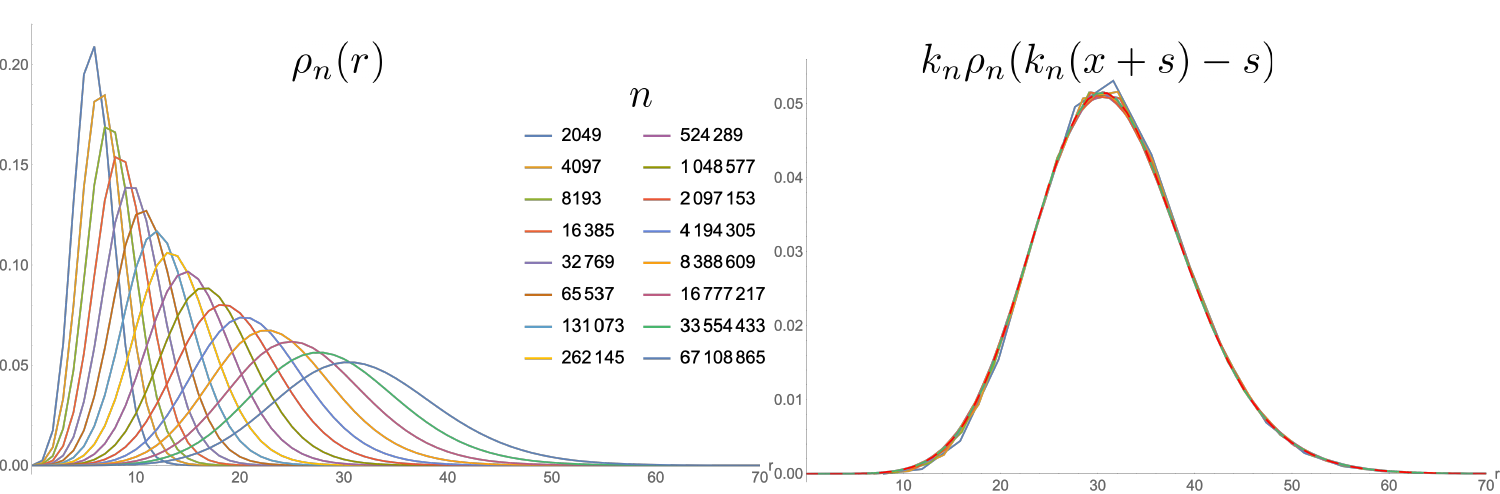}
\caption{Left: Distance histograms $\rho_n(x)$ for $n\in [2^{11}, 2^{26}]$. Right: Example of the fit to $\rho_{n_0}$ with $n_0=2^{19}$ from which the numerical values of $k_n$ are obtained. \label{fig:hist_T3}}
\end{figure}
Comparing these distance histograms with the ones obtained for $D=2$, we observe that they show a consistent scaling. This can be checked by fitting \eqref{eq:num_fit_dh}.
\begin{figure}[H]
\centering
\includegraphics[width=0.5\textwidth]{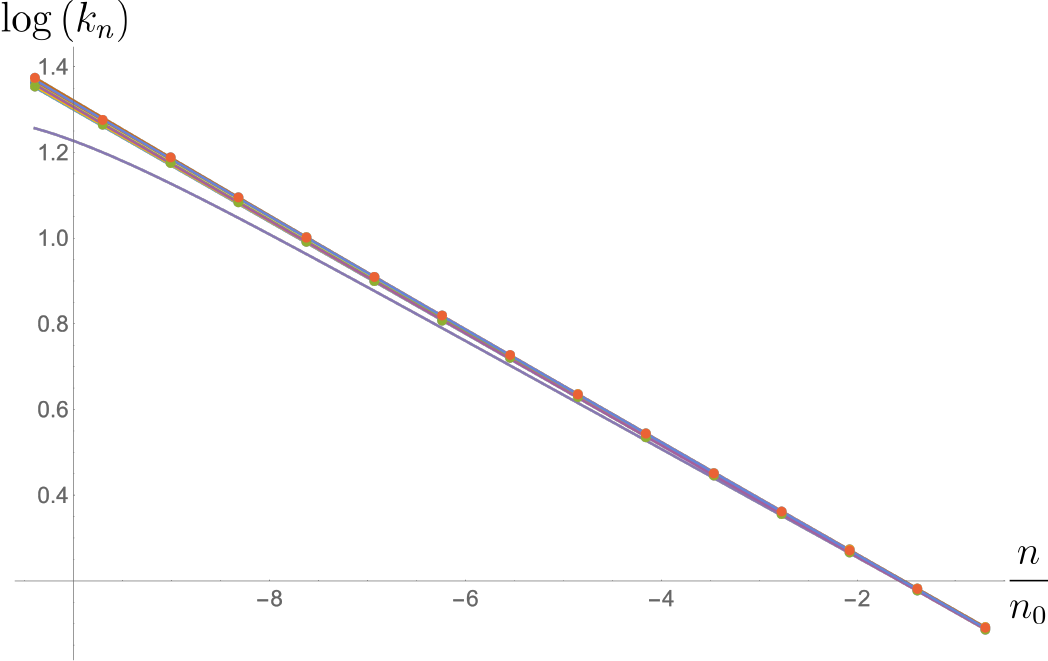}
\caption{Logaritmic plot of fit \eqref{eq:num_fit_dh}.\label{fig:fit_T3}}
\end{figure}

\begin{table}[h]
\centering
\begin{tabular}{|c|c|c|}
\hline
Reference volume & Decile & \textbf{$d_H$} \\
\hline
\multirow{3}{*}{$2^{26}= 67108865$} 
   & 0.75 & $8.10087\pm 0.205694$ \\
   & 0.50 & $7.9087\pm 0.020442$ \\
   & 0.25 & $7.96269\pm 0.177395$ \\
\hline
\multirow{3}{*}{$2^{25}= 33554433$} 
   & 0.75 & $8.04704\pm 0.0924146$ \\
   & 0.50 & $7.89155\pm 0.102166$ \\
   & 0.25 & $7.84954\pm 0.103102$ \\
\hline
\multirow{3}{*}{$2^{24}= 16777217$} 
   & 0.75 & $7.95789\pm 0.364664$ \\
   & 0.50 & $7.92386\pm 0.297647$ \\
   & 0.25 & $7.86777\pm 0.27883$ \\
\hline
\end{tabular}
\caption{Comparison of Hausdorff dimension estimates obtained for $\mathbf{T}^3$ with different numerical schemes given by varying the reference volume $n_0$ and the decile \cite{Fredes2021ModelsOR}.}
\label{tab:dh_tree_d3}
\end{table}
\newpage
\noindent In the case of $\mathbf{R}_n[3]$, we considered the range $n\in [2^{11}, 2^{27}]$. We present an example of the histograms and their fits as well as the logartimic fit.
\begin{figure}[H]
\centering
\includegraphics[width=\textwidth]{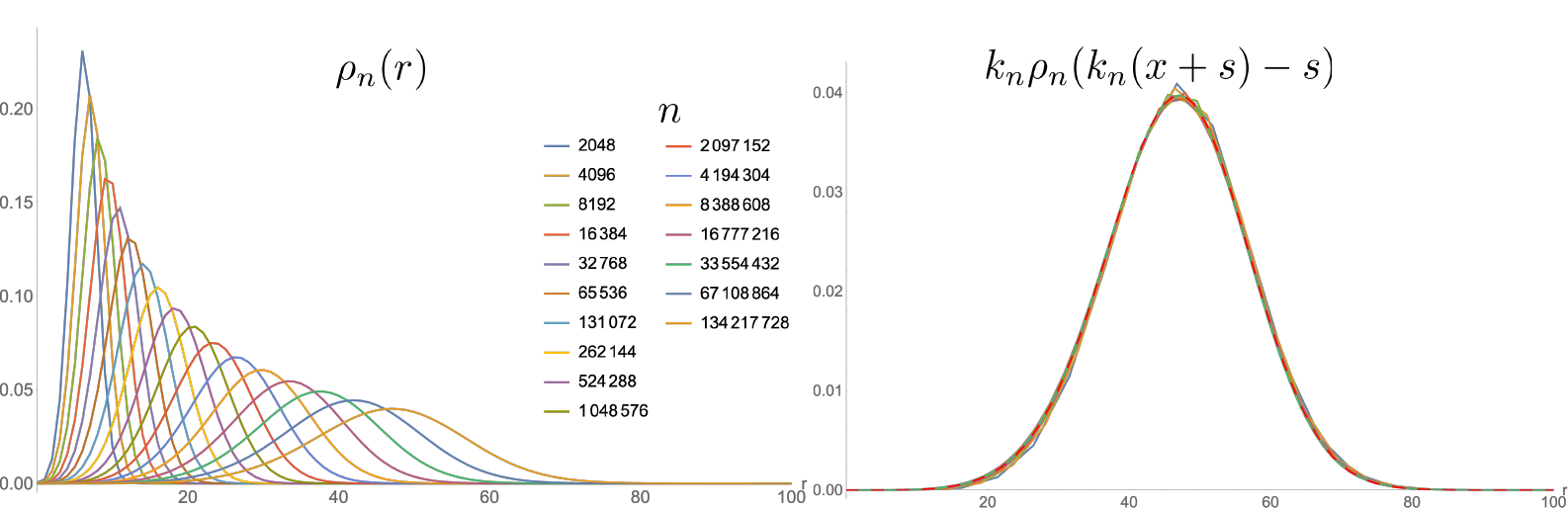}
\caption{Histograms $\rho_n(x)$. Fit to $\rho_{n_0}$ with mean shift and for the numerical estimation of $d^{num}_H$.\label{fig:hist_M3}}
\end{figure}
\begin{figure}[H]
\centering
\includegraphics[width=0.5\textwidth]{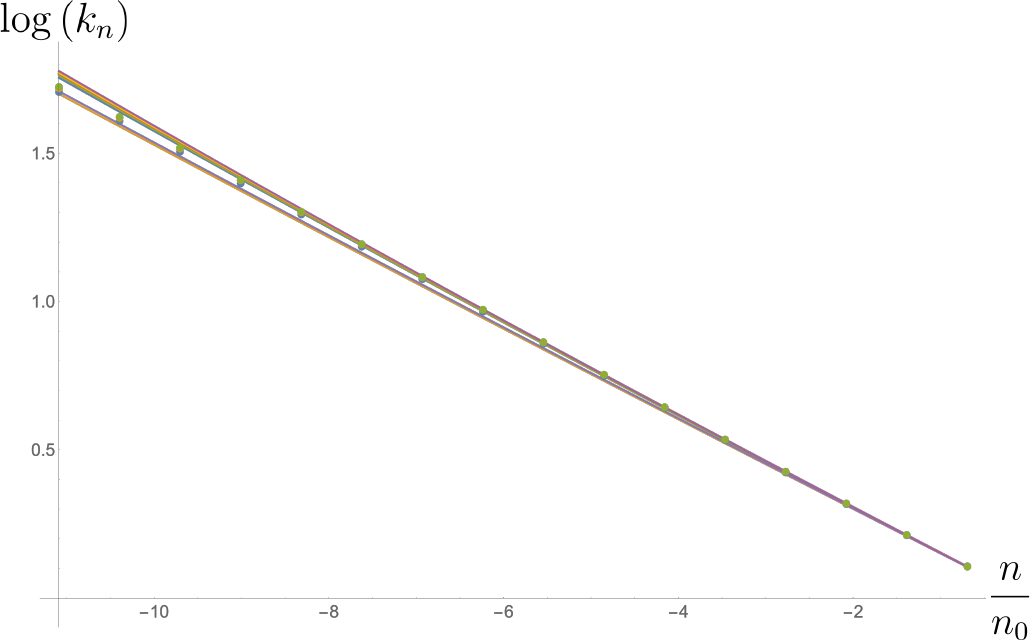}
\caption{Logaritmic plot of fit \eqref{eq:num_fit_dh}.\label{fig:fit_M3}}
\end{figure}
%Given these fits, t
The numerical estimates for the Hausdorff dimension $\dH$ are now given in Table \ref{tab:dh_map_d3}.

\begin{table}[h]
\centering
\begin{tabular}{|c|c|c|}
\hline
Reference volume & Decile & \textbf{$d_H$} \\
\hline
\multirow{3}{*}{$2^{27}= 134217728$} 
   & 0.75 & $7.32657\pm 0.322437$ \\
   & 0.50 & $7.05557\pm 0.106297$ \\
   & 0.25 & $7.12485\pm 0.226403$ \\
\hline
\multirow{3}{*}{$2^{26}= 67108865$} 
   & 0.75 & $7.12307\pm 0.0182644$ \\
   & 0.50 & $7.00843\pm 0.15705$ \\
   & 0.25 & $6.97208\pm 0.188818$ \\
\hline
\multirow{3}{*}{$2^{25}= 33554433$} 
   & 0.75 & $6.79707\pm 0.0387952$ \\
   & 0.50 & $6.75733\pm 0.0512159$ \\
   & 0.25 & $6.69628\pm 0.0447692$ \\
\hline
\multirow{3}{*}{$2^{24}= 16777217$} 
   & 0.75 & $6.64225\pm 0.0844255$ \\
   & 0.50 & $6.62638\pm 0.0547858$ \\
   & 0.25 & $6.61481\pm 0.0727135$ \\
\hline
\end{tabular}
\caption{Comparison of Hausdorff dimension estimates obtained for $\mathbf{R}^3$ with different numerical schemes given by varying the reference volume $n_0$ and the decile \cite{Fredes2021ModelsOR}.}
\label{tab:dh_map_d3}
\end{table}

Given the error expected from analysing the $D=2$-case, we consider a deviation of approximately $8.75\%$ from the theoretical value. That is $\dH= 8 \pm 0.7$. Given that we reached such a threshold and we see a consistent increase in the Hausdorff dimension with increasing volumes, we conclude our numerical estimates are compatible with the conjecture of \cite{Lionni:2019bzb}.
\newpage
\section{Concluding remarks}\label{section:conclusions}

In this paper we have thus implemented a numerical study of the Hausdorff dimension of $D$-random feuilletages introduced in \cite{Lionni:2019bzb}. %By implementing large-scale simulations and finite-size scaling analysis, we were able to estimate the Hausdorff dimension $d_H$ for the first few values of $D$.

For $D=2$, our results reproduce the known value associated with the Brownian map, providing a nontrivial validation of the numerical procedure and of the scaling techniques used. For $D=3$, we obtained results in good agreement with the conjectured value $d_H = 8$, supporting the idea that this model captures a genuinely higher-dimensional universality class of random geometry.

Let us now compare our results with the ones obtained from the mating of trees construction \cite{Budd:2022qlr}. According to \cite{Lionni:2019bzb}, for the $(D=3)$-random feuilletages, the string susceptibility is $\gamma_s=-3/2$. 
On the other hand, according to the numerical results of \cite{Budd:2022qlr}, the string susceptibility exponent $\gamma_s=-3/2$ corresponds approximately to the mating of three trees with correlation $-\cos(0.7)\pi$. Moreover, we know that the first pair of trees are correlated according to the law of the label process of uniform quadrangulations.
Thus, this implies that $-\cos(2\pi/3)=1/2$. However, when using this value and the fit in \cite{Budd:2022qlr} for isosceles regions, we get that there is no real solution for the angle $\beta$, {\it i.e.} there is no mated-CRT map in the isosceles region for which both of the trees are correlated according to uniform quadrangulations and for whom the Hausdorff dimension is equal to $8$.
Therefore, this argument indicates that the $(D=3)$-feuilletage most likely does not belong to the $D=3$ mating of trees universality classes.

\medskip

Let us end this paper by emphasizing that our findings provide numerical evidence suggesting that the $D$th random feuilletages has a suitable scaling-limit and it constitutes a natural  candidate for a higher-dimensional generalization of the Brownian map as a metric space. From the theoretical physics point of view, this opens the way to an exploration of the continuum limit of these discrete geometries and of their potential connections to models of quantum gravity, where such scale-invariant geometries could play a fundamental role.

\section*{Acknowledgments}
We warmly acknowledge Jean-François Marckert for sharing the code used to produce the discrete feuilletages, as well as for several discussions at various stages of this project. This work was supported by the ANR-20-CE48-0018 “3DMaps” grant.

\bibliographystyle{utphys}
\bibliography{references}

\end{document}